\documentclass{article}

\usepackage{arxiv}

\usepackage[utf8]{inputenc} 
\usepackage[T1]{fontenc}    
\usepackage{hyperref}       
\usepackage{url}            
\usepackage{booktabs}       
\usepackage{amsfonts}       
\usepackage{nicefrac}       
\usepackage{microtype}      
\usepackage{lipsum}		
\usepackage{graphicx}
\usepackage{doi}
\usepackage{authblk}
\usepackage{here}
\usepackage{indentfirst}

\title{PlaNet-S: Automatic Semantic Segmentation of Placenta}

\date{} 					

\author[1, $\dag$]{Shinnosuke Yamamoto MS}
\author[1, $\dag$]{Isso Saito MS}
\author[1, 2]{Eichi Takaya ME}
\author[3, 4]{Ayaka Harigai MD}
\author[3]{\\Tomomi Sato MD/PhD}
\author[1]{Tomoya Kobayashi RT/PhD}
\author[4]{\\Kei Takase MD/PhD}
\author[1, 2, 3, $*$]{Takuya Ueda MD/PhD}

\affil[1]{Department of Clinical Imaging, Tohoku University Graduate School of Medicine, Japan}
\affil[2]{AI Lab, Tohoku University Hospital, Japan}
\affil[3]{Department of Diagnostic Radiology, Tohoku University Hospital}
\affil[4]{Department of Diagnostic Radiology, Tohoku University Graduate School of Medicine}

\renewcommand{\shorttitle}{PlaNet-S}

\begin{document}
\raggedright
\setlength{\parindent}{12pt}
\maketitle

\def\thefootnote{$\dag$}\footnotetext{Shinnosuke Yamamoto, MS and Isso Saito, MS equally contributed to this work.}\def\thefootnote{\arabic{footnote}}
\def\thefootnote{$*$}\footnotetext{Correspondence to: to Takuya Ueda, MD, PhD, Department of Clinical Imaging, Tohoku University Graduate School of Medicine, Seiryo-machi 2-1, Aoba-ku, Sendai, Miyagi 980-8575, Japan; Tel: +81-22-717-7481; Fax: +81-22-717-7944; Email:takuya.ueda.d3@tohoku.ac.jp}\def\thefootnote{\arabic{footnote}}

\begin{abstract}
\textbf{Purpose:} To develop a fully automated semantic placenta segmentation model that integrates the U-Net and SegNeXt architectures through ensemble learning. 

\textbf{Methods:} A total of 218 pregnant women with suspected placental anomalies who underwent magnetic resonance imaging (MRI) were enrolled, yielding 1090 annotated images for developing a deep learning model for placental segmentation. The images were standardized and divided into training and test sets. The performance of PlaNet-S, which integrates U-Net and SegNeXt within an ensemble framework, was assessed using Intersection over Union (IoU) and counting connected components (CCC) against the U-Net model.  

\textbf{Results:} PlaNet-S had significantly higher IoU (0.73, SD = 0.13) than that of U-Net (0.78, SD = 0.10) (p<0.01). The CCC for PlaNet-S was significantly higher than that for U-Net (p<0.01), matching the ground truth in 86.0\% and 56.7\% of the cases, respectively.

\textbf{Conclusion:} PlaNet-S performed better than the traditional U-Net in placental segmentation tasks. This model addresses the challenges of time-consuming physician-assisted manual segmentation and offers the potential for diverse applications in placental imaging analyses.
\end{abstract}

\keywords{Deep learning \and Placenta \and Magnetic resonance imaging \and Automatic semantic segmentation \and Vision transformer}

\section{Introduction}
As a critical link between the mother and fetus, the placenta is essential for both pregnancy success and fetal growth. Abnormalities in placental function may pose severe risks to both maternal and fetal health, potentially resulting in irreversible harm or life-threatening situations \cite{Silver2015}. Conditions such as placenta previa \cite{Gude2004}, placenta accreta spectrum (PAS) \cite{Booker2019, Bailit2015, Jauniaux2019}, and uteroplacental insufficiency \cite{Nakao2023} are among the most common reasons for imaging evaluations during pregnancy. Assessment of the placenta is crucial for understanding placental structure, function, and development and to identify strategies to optimize pregnancy outcome \cite{Gagnon2003, Warshak2010, Leyendecker2012}.

In placental imaging, the location, shape, and volume of the placenta are important factors for identifying abnormalities. Ultrasonography is the primary imaging modality for monitoring fetal health and development. Magnetic resonance imaging (MRI)-based evaluation of the placenta provides more detailed anatomical and functional information, especially for complex cases or posteriorly located placentas where ultrasound application is limited \cite{Leyendecker2012, Do2019}. Several studies have reported the usefulness of placental segmentation using MRI for monitoring conditions that can lead to pregnancy and birth complications, such as PAS, fetal growth restriction, and potential intrauterine fetal demise \cite{Anquez2013, Collins2013, Wang2016}. Despite its utility, manual segmentation in MRI is time-consuming and exhibits high inter- and intra-observer variability \cite{Clarke1995}.

Deep learning (DL) -based approaches, especially those based on the convolutional neural network (CNN) architecture, have demonstrated a strong capability for the fast and automatic segmentation of medical images with high accuracy \cite{Shahedi2021}. U-net is a representative CNN architecture specifically designed for image-segmentation tasks \cite{Ronneberger2015}. Although several researchers have proposed automatic segmentation methods for the human placenta using the U-Net architecture, U-Net-based segmentation models still require improvement in terms of robustness and generalizability \cite{Han2019, Shahedi2022}. Recently, the Vision Transformer (ViT) was introduced as a novel architecture for DL-based image analysis \cite{Dosovitskiy2021}. ViT was adapted from the Transformer model originally designed for natural language processing \cite{Vaswani2017}. Whereas CNN-based models predominantly rely on local textures in images \cite{Baker2018, Geirhos2019}, ViT-based models emphasize the global shape of objects for image recognition \cite{Tuli2021}. Although ViT-based models have the potential to outperform CNN-based DL models in the field of semantic segmentation when trained on large datasets \cite{Dosovitskiy2021}, their performance is heavily dependent on the amount of data \cite{Willemink2022}. As large-scale data are not always available in medical research, these transformers may not achieve optimal performance in such medical contexts. 

SegNeXt (a simple convolutional network architecture for semantic segmentation), proposed by Guo et al., is a streamlined CNN architecture designed specifically for semantic segmentation \cite{Guo2022}. To address the abovementioned limitations of ViT-based models, SegNeXt was designed to incorporate multiscale convolutional attention to encode contextual information, emphasizing cost-effective convolutional operations. By emulating the ViT mechanism and embedding it within a CNN framework, SegNeXt enhances its performance while concurrently optimizing computational expenses \cite{Guo2022}.

This study aimed to develop a fully automated semantic placenta segmentation model by integrating ensemble learning techniques with the U-Net and SegNeXt architectures to enhance the reliability and precision of placental tissue delineation in MRI.

\section{Materials \& Methods}
\subsection{Patient Selection and Enrollment}
In this study, we included placental MRIs scans performed at our hospital between January 2004 and December 2021. Single-shot T2 weighted sequence images from 218 patients with suspected placental abnormalities were analyzed. Ninety-five patients were diagnosed with placenta previa, fifty-five with placenta accreta spectrum, and fifty-two with both.

\subsection{MRI Equipment and Parameters}
MRI in this study was performed using the following four MRI scanners: 1.5T-Achieva (Philips Medical Systems, Best, the Netherlands), 3T-MAGNETOM Vida (Siemens Healthcare, Erlangen, Germany), 1.5T-MAGNETOM VISION plus (Siemens Healthcare, Erlangen, Germany), and 3T-Vantage Titan 3T (Cannon Medical Systems, Tokyo, Japan). T2-weighted imaging of placenta was obtained using single-shot fast spin-echo T2-weighted sequences with repetition time (TR), $\infty$ ms; echo time (TE), 64–120 ms; acquisition matrix, 178×224-287×384; field of view (FOV), 263×300-420×420; and slice thickness, 4-10 mm.

\subsection{Image Selection and Annotation}
A radiologist with 12-y expertise in gynecological imaging selected five consecutive images in which the placenta was clearly visible. Consequently, our dataset comprised 1090 images from 218 patients. For each image, a radiologist annotated the placental region using a home-made annotation device.

\subsection{Preprocessing and Image Dataset}
To standardize the resolution of all MR images to 256×256 pixels, each image was resized based on its longest dimension, that is, height or width. The images were scaled such that the longest dimension was 256 pixels while preserving the original aspect ratio. For images that did not form a perfect square after resizing, the original image was centrally placed and the shorter dimension was symmetrically padded with black pixels, ensuring the desired 256×256 pixel size. Our original dataset, comprising 1090 images, was divided into training and test datasets in an 8:2 ratio, resulting in 875 and 215 images for the training and test datasets, respectively. The dataset was divided based on the patients to avoid image overlaps from the same patient between the training and test datasets, thus ensuring that all five MR images from a single patient were allocated to one of the training and test datasets.

\subsection{Model Architecture of Placental Segmentation Network (PlaNet-S)}
In this study, a specialized DL model named the Placental Segmentation Network (PlaNet-S) is proposed for performing accurate segmentation of the placenta in MR imaging. Figure 1 illustrates the structural design and ensemble learning approach inherent in the PlaNet-S architecture. PlaNet-S employs ensemble learning techniques that integrate the U-Net and SegNeXt architectures (Figure 1) \cite{Ronneberger2015, Guo2022}.

Figure 2 presents part of the architectural design of the PlaNet-S model, which is deeply rooted in the SegNeXt variant. The depicted section of the architecture unfolds across multiple stages, starting from Stage 1 and extending to Stage 4. Each stage shows a distinct iteration of the SegNeXt-S framework marked by varying the numerical values that potentially represent the number of feature channels or neurons. Notably, the design also incorporates the SegNeXt-S encoder, which is seamlessly paired with a straightforward multilayer perceptron in the decoder role, drawing parallels to the configurations observed in models such as SegFormer \cite{Xie2021}.

The loss function (L) for PlaNet-S is defined in Equation (1). $L_{BCE}$ represents binary cross-entropy. $L_{IoU}$ is defined by Equation (2) using Intersection over Union (IoU): The IoU is the ratio of the overlap area between the predicted and true bounding boxes and their union area, which can be used to quantify the accuracy of the segmentation \cite{Fu2019}. Higher IoU scores indicated more accurate predictions of the true bounding boxes.

\begin{equation}
L = L_{BCE} + L_{IoU}
\end{equation}

\begin{equation}
L_{IoU} = 1 - IoU
\end{equation}

\begin{figure}[htbp]
  \centering
  \includegraphics[width=170mm]{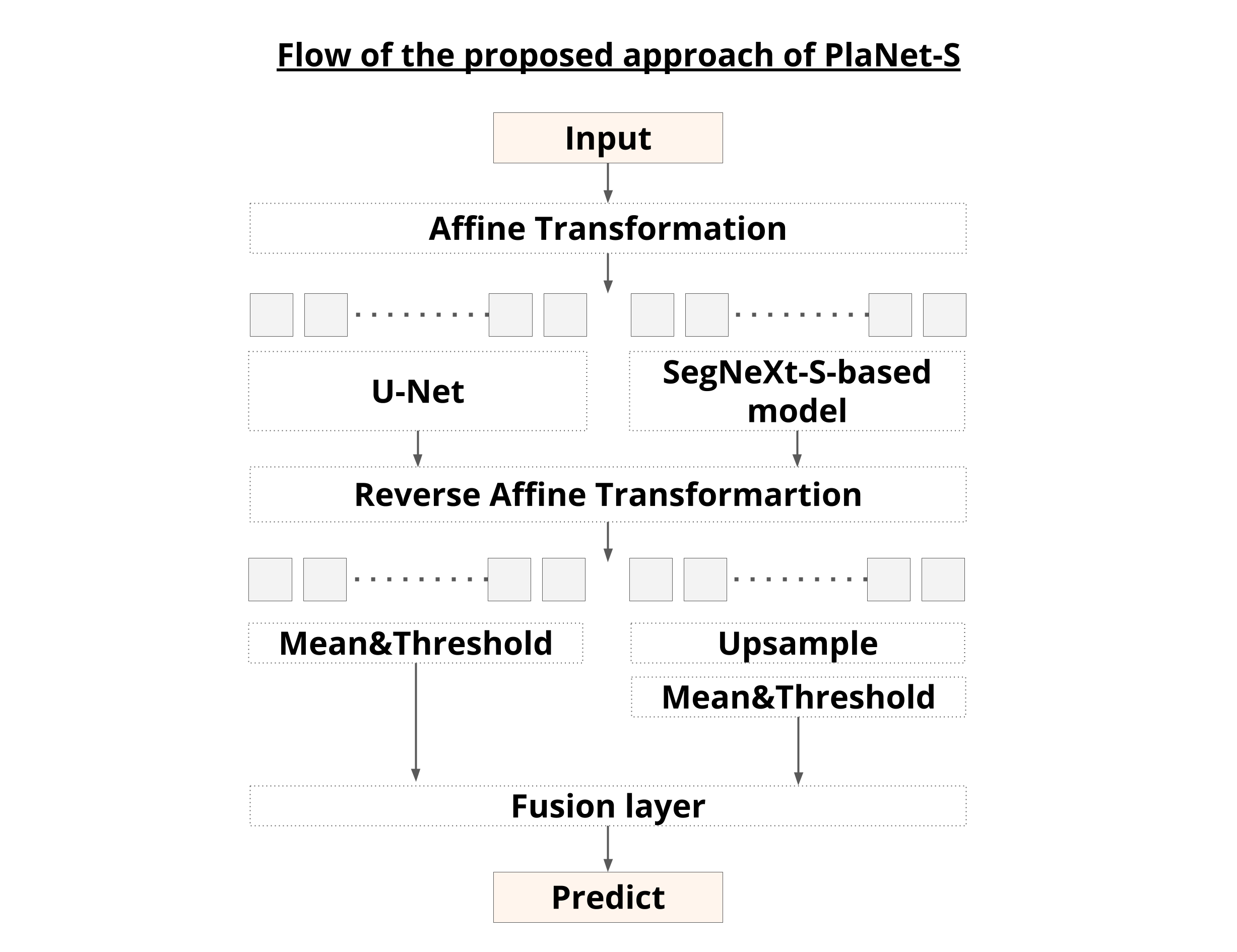}
  \label{fig:fig1}
\begin{flushleft}
\textbf{\centerline{Figure 1: Flow of the proposed approach of PlaNet-S}} 

The Placental Segmentation Network (PlaNet-S) is a specialized deep learning approach tailored for precise placental segmentation in MR imaging. The proposed model employs ensemble learning by integrating U-Net and SegNeXt. The process begins with an affine transformation applied to the input to create 200 variants. The 200 variants were distributed across two distinct models, each generating 100 outputs. After processing, the model predictions were reverted to their original spatial alignment using the inverse of the initial transformation. Predictions that met a specific threshold were combined to yield the final output from the union of the model results.
\end{flushleft}
\end{figure}

\begin{figure}[htbp]
  \centering
  \includegraphics[width=170mm]{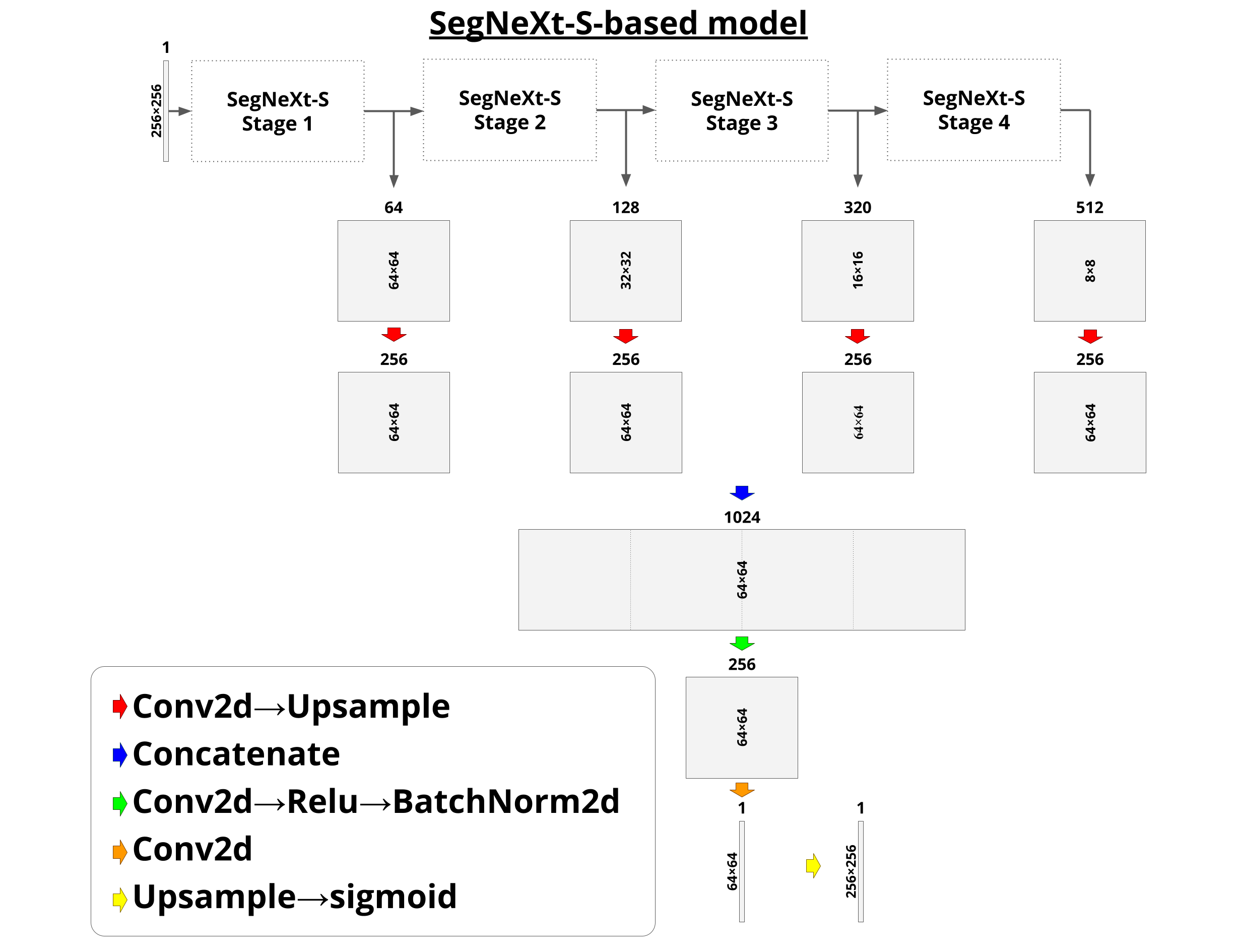}
  \label{fig:fig2}
\begin{flushleft}
\textbf{\centerline{Figure 2: SegNeXt-S-based model}}

The structure progresses through stages 1–4, with each stage depicting a unique facet of the SegNeXt-S framework. The numerical annotations indicated varying complexities, possibly corresponding to the number of feature channels or neurons. The feature maps generated using the SegNeXt Encoder were transitioned to 256 channels and subsequently fused to form a 64×64×1024 feature volume. After two convolutional stages, the data were up-sampled, culminating in a final resolution of 256×256×1 pixels. The SegNeXt-S encoder is adeptly matched with a multilayer perceptron decoder, mirroring the design patterns in models such as SegFormer.
\end{flushleft}
\end{figure}

\subsection{Training step}
During the PlaNet-S training step, an affine transformation was utilized for data augmentation. Three elements of affine transformation were used: rotation, ranging from -45 to +45 degrees; shift, ranging from 0 to 1 in each of the (x, y) axes; and scaling, ranging from 0.8 to 1, with all parameters being randomly applied.

PlaNet-S was implemented using an NVIDIA Quadro RTX 8000 machine. The operating system was Ubuntu 18.04.5 long-term support (LTS): Xenial Xerus. All analyses were performed using Python, version 3.7.3 (Python Software Foundation, http://www. python. org). Pytorch version 1.9.5 was used as the deep-learning framework.

\subsection{Test-time augmentation(TTA)}
Test-Time Augmentation (TTA) was also applied for data augmentation during both the validation and testing phases. Although data augmentation is typically employed during the training step, TTA adopts a similar approach to improve the robustness toward the test and validation steps \cite{Wang2019, Shorten2019}. Mathematically, given an input image $x$, transformations $\phi_1$, $\phi_2$,… are applied to produce a set of transformed images $x_1$, $x_2$,… . For each transformed image $x_i$, model $M$ provides the prediction $p_i$. The final prediction, $P$ is then computed by averaging the individual predictions.
In terms of formula:

\begin{equation}
p_{i} = M(\phi_{i}(x))
\end{equation}

\begin{equation}
P = \frac{1}{N}\sum_{i=1}^{N}p_{i}
\end{equation}

\subsection{Validation step}
To mitigate the potential for overfitting the DL model, 5-fold cross-validation was conducted. From each fold, the model that achieved the IoU score was selected as the definitive model. To enhance the robustness during validation, TTA was applied to the validation step.

\subsection{Test step}
For the segmentation task, the final prediction of PlaNet-S was assessed using a test set comprising 43 cases, each containing 5 images, yielding 215 images. The IoU was then calculated for each set of predictions. The TTA was applied during the test step to enhance the robustness of the model.

\subsection{Comparison of model performance of PlaNet-S and traditional U-Net}
To assess the performance of PlaNet-S, its predictions were compared with those of the traditional segmentation model, U-Net, using the same dataset. The IoU and Counts of Connected Components (CCC) were employed as criteria for performance assessment. The IoU, a well-established metric for assessing segmentation accuracy, quantifies the overlap between the predicted and true bounding boxes relative to their combined area. A higher IoU value typically indicates a more accurate depiction of placental segmentation. Additionally, we adopted as an auxiliary measure of segmentation precision. The CCC delineates distinct segmented regions in an image. Each connected component epitomizes a cluster of pixels or voxels either directly or indirectly, denoting discrete entities within the segmented image. Segmentation was considered more accurate when no difference was observed between the CCC calculated from the ground-truth annotation and that derived from the segmentation of the deep learning model. A paired t-test was used to assess the statistical significance of the difference in the IoU between PlaNet-S and U-Net. p<0.01 is considered as statistically significant. The Wilcoxon signed-rank test was used to assess the statistical significance of the differences in the CCC between PlaNet-S and U-Net, excluding pairs with identical results. p<0.01 is considered as statistically significant.

\section{Results}
Figure 3 shows box plots depicting the IoU for PlaNet-S and U-Net, which showed means of 0.73 (SD = 0.13) and 0.78 (SD = 0.10) respectively. The IoU for PlaNet-S is significantly higher (p<0.01).

\begin{figure}[H]
  \centering
  \includegraphics[width=80mm]{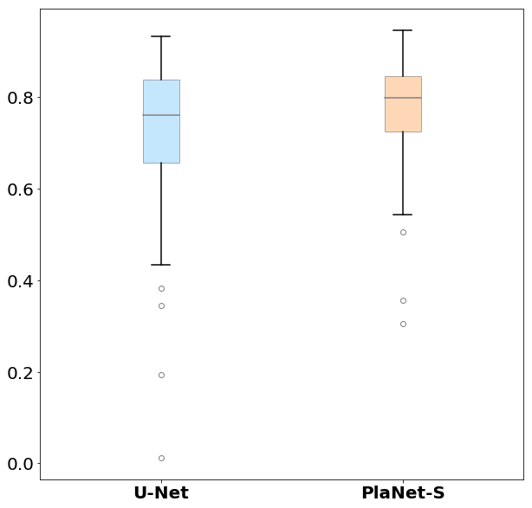}
  \label{fig:fig3}
\begin{flushleft}
\textbf{\centerline{Figure 3: IoU (Intersection over Union) Box plot for PlaNet-S vs. U-Net}} 

The distribution of the prediction accuracy, represented by the IoU for PlaNet-S compared with U-Net, is depicted in a box plot format. PlaNet-S and U-Net showed means of 0.73 (SD = 0.13) and 0.78 (SD = 0.10) of IoUs, respectively. The IoU for PlaNet-S is significantly higher (p<0.01).
\end{flushleft}
\end{figure}

\begin{figure}[H]
  \centering
  \includegraphics[width=80mm]{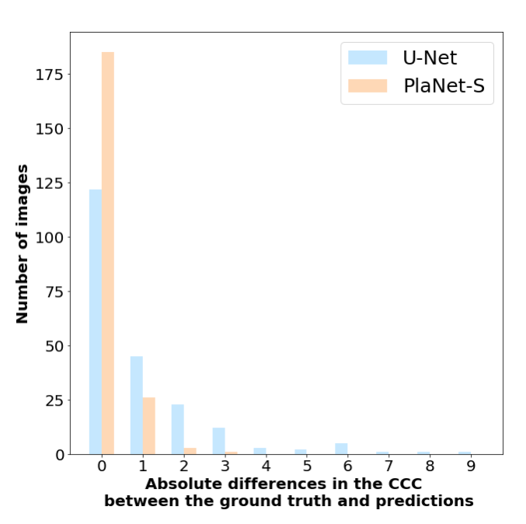}
  \label{fig:fig4}
\begin{flushleft}
\textbf{
    \centerline{Figure 4: Bar plot comparing the absolute differences in the Counts of Connected Components}
    \centerline{between ground truth and predictions for PlaNet-S and U-Net.}
}
The histogram illustrates the distribution of the counts of connected components (CCC) across 215 segmentation outcomes from both models. The X-axis represents the absolute differences in the CCC between the ground truth annotations and segmentations performed by the DL models (PlaNet-S and U-Net), whereas the y-axis indicates the cumulative number of patients corresponding to those differences. The CCCs for PlaNet-S were significantly lower than those for PlaNet-S (p<0.01).
\end{flushleft}
\end{figure}

Figure 4 shows a bar plot depicting the absolute difference between the ground-truth number of Connected Components and the inferred Connected Components calculated from the results of PlaNet-S and U-Net. PlaNet-S and U-Net achieved segmentation with an absolute difference of zero in the number of connected components for 185 and 122 test images, representing 86.0\% and 56.7\% of the dataset, respectively. The CCCs of PlaNet-S were significantly smaller than those of U-Net (p<0.01).

Figure 5 shows a comparison of the segmentation outcomes of PlaNet-S and U-Net across the four illustrative cases. In Case 1, both models demonstrated congruence with the ground truth, underscoring their potential for accurate segmentation. Case 2 highlights the superior performance of PlaNet-S, which remained consistent with the ground truth, whereas U-Net exhibited a notable deviation. By contrast, Case 3 offers an instance in which PlaNet-S diverges significantly from the ground truth, whereas U-Net maintains the alignment. Finally, Case 4 provides an example in which neither model achieves concordance with the ground truth, indicating areas for further refinement.

\begin{figure}[H]
  \centering
  \includegraphics[width=140mm]{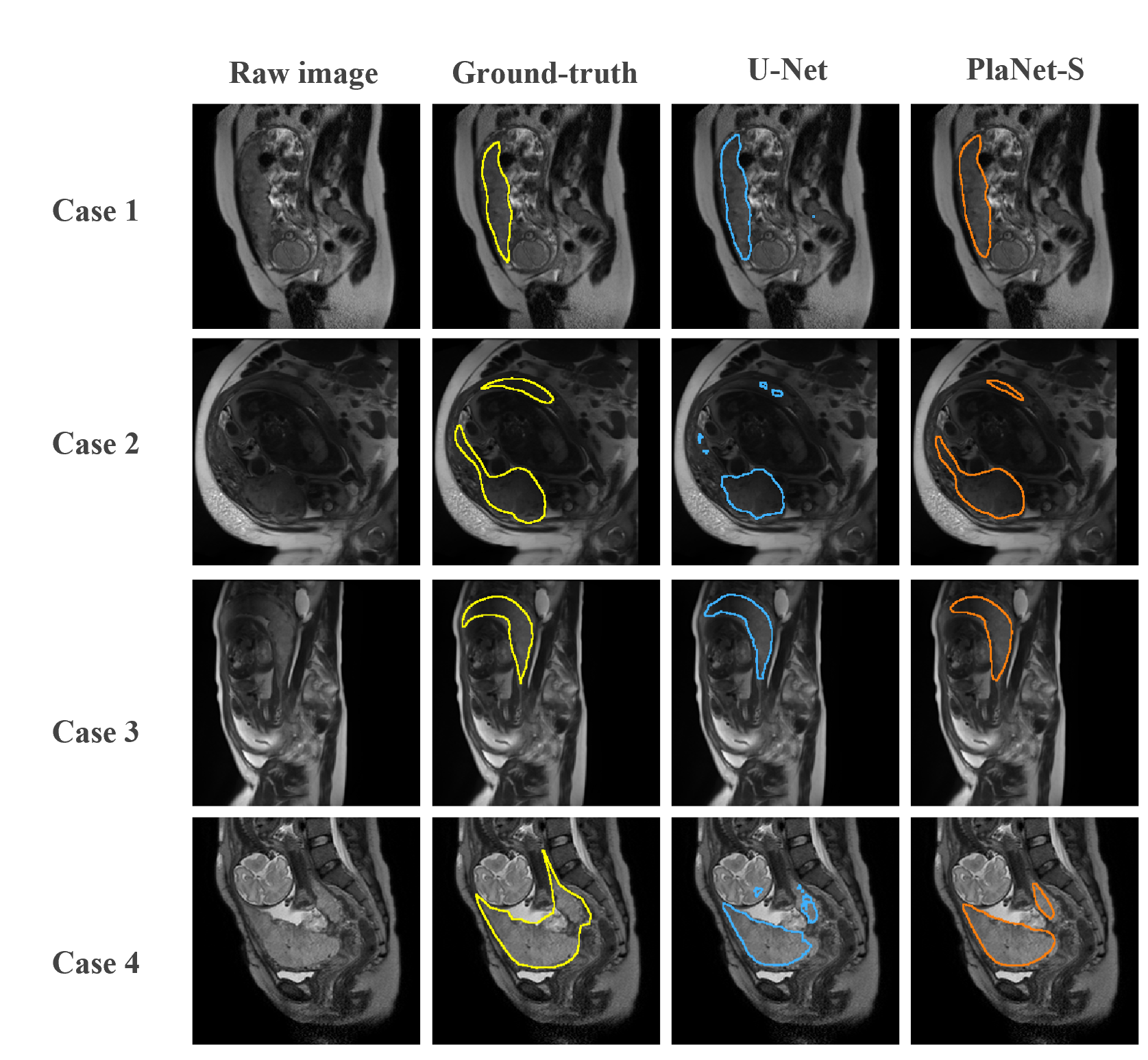}
  \label{fig:fig5}
\begin{flushleft}
\textbf{\centerline{Figure 5: Comparative Examples of Segmentation Results for PlaNet-S vs. U-Net}} 

Figure 5 shows a comparison of the segmentation outcomes of PlaNet-S and U-Net across the four illustrative cases. In Case 1, both models demonstrated congruence with the ground truth, underscoring their potential for accurate segmentation. Case 2 highlights the superior performance of PlaNet-S, which remained consistent with the ground truth, whereas U-Net exhibited a notable deviation. In contrast, Case 3 offers an instance in which PlaNet-S diverges significantly from the ground truth, whereas U-Net maintains the alignment. Finally, Case 4 provides an example in which neither model achieves concordance with the ground truth, indicating areas for further refinement.
\end{flushleft}
\end{figure}

\section{Discussion}
Our proposed DL model, PlaNet-S, which incorporates a heterogeneous ensemble learning approach combining both SegNeXt and U-Net, demonstrated superior segmentation results compared to conventional U-Net alone. Whereas SegNeXt emphasizes the global shapes of objects during image recognition, U-Net primarily leverages local textures in images. As the overall morphology of the placenta is typically represented as a relatively cohesive structure, appearing as one or a few consolidated masses, and its textures are marked by high-intensity signals on T2WI, the ensemble learning approach combining SegNeXt and U-Net might be successful in capturing both these aspects. 

In our study, TTA was implemented during the test process to leverage its inherent strengths and enhance the accuracy of placenta segmentation. The TTA is renowned for its ability to provide robust and stable predictions by incorporating multiple augmented views of the same image. Such augmentation-driven robustness can be particularly beneficial in medical imaging, where slight variations or perspectives in imagery can affect a model's decision. TTA contributes to improving accuracy, mitigating overfitting, and aiding in better generalization of unseen data. These characteristics of TTA also play a significant role in showing a more cohesive and consistent segmentation, which might improve the CCC indicators. Consequently, TTA contributed to the improved performance of our placenta segmentation model.

The development of PlaNet-S led to significant advancements in medical imaging, potentially resolving the traditional time-consuming and specialist-dependent processes of placental segmentation. The automation provided by PlaNet-S streamlines workflow and reduces the need for specialized expertise. Future research should pivot toward a fully automated diagnostic AI capable of identifying perinatal conditions such as placenta accreta, preeclampsia, and fetal growth restriction. Combined with deep-learning technologies for classifying placental diseases, PlaNet-S has the potential to form a comprehensive diagnostic model for placental disorders. Furthermore, the principles underpinning PlaNet-S can be extended to the segmentation and diagnosis of various other organs, thereby broadening its impact on medical diagnostics.

This study has two limitations. First, the dataset was obtained from a single facility. Sourcing data from a single facility may limit the generalizability of the model, introduce bias from specific equipment or demographics, and increase the risk of overfitting the dataset. Future studies should prioritize diversifying the data sources and employing advanced data augmentation techniques during training to enhance placental segmentation and ensure broader applicability. Second, PlaNet-S has the limitation that it may not segment well when the placenta is thin and elongated, although this is less frequent than when using U-Net.

\section{Conclusion}
Utilizing ensemble learning with SegNeXt and U-Net, PlaNet-S showed a higher performance on the segmentation task of the placenta than the traditional U-Net. This model addresses the challenges of time-consuming physician-assisted manual segmentation and offers the potential for diverse applications in placental imaging analyses.

\section*{Acknowledgments}
The authors thank Dr. Masatoshi Saito and the other members of the Department of Gynecology and Obstetrics at Tohoku University Graduate School of Medicine, as well as the participants of our study, for their contributions.

\section*{Ethic}
This retrospective study was approved by the institutional review board (IRB no: 2022-1-114-1) and the requirement for informed consent from patients was waived.

\section*{Data Availability}
Data from this study are not publicly available and are restricted to protect patient privacy in accordance with the institutional review board guidelines. The data were securely stored under controlled access at Tohoku University Hospital. Requests for data access may be submitted to the corresponding author and considered by the IRB on a case-by-case basis, subject to an approved research proposal that meets the requisite criteria for ethical handling and privacy protection.

\section*{Funding}
This work was supported by the JST (CREST Grant No. JPMJCR15D1); MEXT/JSPS WISE Program: Advanced Graduate Program for Future Medicine and Health Care, Tohoku University, Japan; and JSPS KAKENHI Grant Number JP20K16687.

\section*{Conflict of interest}
The author(s) have no conflicts of interest to disclose.

\bibliographystyle{unsrt}  
\bibliography{PlaNet_S}  

\end{document}